\documentclass[aip,graphicx]{revtex4-1}

\draft 

\usepackage{graphicx}

\begin{document}


\title{Slow-Fast Brain-Computer Interfaces: Preventing Neuroadaptive Overfitting in AI-Mediated Neural Interfaces} 



\author{Aarthy Nagarajan}
\affiliation{Department of Computer Science and Engineering, University of Notre Dame, Notre Dame, IN, USA}


\date{\today}

\begin{abstract}
Artificial intelligence (AI) is transforming brain-computer interfaces (BCIs) from task-specific neural decoders into adaptive systems that complete language, smooth movement, regulate rehabilitation support and adjust stimulation. These capabilities can increase speed, fluency, usability and clinical reach, yet conventional performance metrics may overlook losses in intent fidelity, authorship, agency, therapeutic challenge and durable clinical benefit. I define neuroadaptive overfitting as a closed-loop failure mode in which an AI-mediated BCI becomes over-optimized to short-term proxies of success, including reduced effort, rapid acceptance, lower workload or smooth task completion, while drifting from the user’s durable goals. I then propose Slow–Fast BCI, a framework for pacing AI assistance according to decoder evidence, uncertainty, contextual and clinical stakes, fatigue, and user- or clinician-defined goals. The framework distinguishes fast assistance when intent is clear and stakes are low, guarded assistance under uncertainty and slow assistance when misalignment could compromise safety, agency, authorship, motor learning or therapeutic value. Across communication, motor-control, neurorehabilitation and closed-loop neuromodulation applications, I outline corresponding safeguards and evaluation measures. This Perspective argues that AI-mediated BCIs should be evaluated not only by decoding accuracy and task performance, but also by how AI assistance is deployed: when systems act autonomously, seek confirmation, preserve user effort or return control to the user.
\end{abstract}


\maketitle 

\section{Introduction}

Brain-computer interfaces are moving from task-specific neural decoders toward AI-mediated systems. Classical BCIs translate neural activity into communication or control outputs \cite{Wolpaw2002Brain-computerControl}, and adaptive systems use machine learning, decoder adaptation and user feedback to improve calibration or usability \cite{Vidaurre2011}. BCI performance also reflects learning on the user side of the closed loop, with recent work showing that users acquire control through structured changes in neural activity during BCI learning \cite{Busch2026HumanGeometry}. What is changing is the addition of AI assistance layers that use language priors, autocomplete, adaptive control and emerging generative models to complete, rank, smooth or act on partially decoded user intent \cite{Willett2023ANeuroprosthesis,Metzger2023AControl,Lee2025BraincomputerCopilots}. These layers can improve communication, reduce fatigue and extend clinical reach, while also shaping how the final utterance, movement or intervention is completed.

These changes complicate BCI evaluation because improvements in speed, smoothness or ease of use need not imply closer alignment with the user’s underlying goals. Language assistance, especially as systems move from constrained decoding toward richer autocomplete or generative support, can increase fluency, creating a risk that model-driven completions diverge from intended meaning or perceived authorship \cite{Willett2023ANeuroprosthesis,Metzger2023AControl}. Shared autonomy can improve movement trajectories, while increasing the difficulty of disentangling user control from autonomous system contribution \cite{Lee2025BraincomputerCopilots,Davidoff2020AgencyInterfaces.,Haag2025EthicalReview, M.Zhangetal.2026AnTraining}. In neurorehabilitation, assistance can reduce frustration and facilitate task completion, yet excessive assistance risks diminishing the effort and error experience needed for motor learning \cite{Jin2024Electroencephalogram-basedReview,Marchal-Crespo2009ReviewInjury}. Closed-loop neuromodulation can optimize short-term symptom-related proxies, creating a risk that immediate gains diverge from longer-term safety or clinical durability \cite{Haag2025EthicalReview,Amodei2016ConcreteSafety, Oehrn2024ChronicTrial}.

I use the term \emph{neuroadaptive overfitting} to describe a closed-loop condition in which adaptation increasingly favors short-term proxies of user state or task success—such as reduced effort, rapid acceptance, lower workload or smooth performance—while degrading higher-order goals such as agency, intent fidelity, therapeutic challenge or long-term clinical benefit.
Operationally, neuroadaptive overfitting is indicated when increasing or adapting AI assistance improves a proximal performance metric while degrading one or more prespecified user-centred or clinical outcomes. A necessary condition is that the assistance policy, or its effective behavior, changes over repeated interaction in response to neural, behavioral, contextual or acceptance signals, such that the closed loop increasingly optimizes a proxy. A fixed but overly aggressive assistance policy would therefore constitute over-assistance rather than neuroadaptive overfitting. This divergence should reflect the assistance policy rather than decoder error alone, and the relevant proximal and durable outcomes will vary across BCI applications. Neuroadaptive overfitting may appear as semantic drift in communication BCIs, reduced agency in motor-control BCIs, or maladaptive assistance in rehabilitation, and may extend to adjacent closed-loop neural interfaces such as neuromodulation. The term draws on two related ideas: 1) overfitting, in which a system tunes too closely to limited or misleading signals rather than to the intended objective, and 2) sycophancy, in which systems optimized with human feedback may learn to please or affirm users at the expense of calibration or truthfulness \cite{Sharma2024TowardsModels}. In BCIs, however, neuroadaptive overfitting is not ordinary offline model overfitting. It therefore arises during ongoing human-machine interaction when adaptation increasingly favors convenient neural, behavioural or contextual proxies over durable user goals \cite{Davidoff2020AgencyInterfaces.,Haag2025EthicalReview,Felton2012MentalTraining,Kubler2014TheApplications}. If these proxies become optimization targets, AI-mediated BCIs risk favouring what is easy to infer or immediately rewarding over what is faithful, effort-preserving or clinically valuable \cite{Marchal-Crespo2009ReviewInjury,Lakshminarayanan2017SimpleEnsembles,Amodei2016ConcreteSafety}.

Recent co-adaptation studies show that changes in decoder behavior can alter user effort even when task performance changes little \cite{Madduri2026ComputationalInterfaces}. Neuroadaptive overfitting extends this concern from user--decoder co-adaptation to adaptive AI assistance policies, asking whether improvements in proximal performance conceal deterioration in prespecified durable user- or clinician-centered outcomes.

Neuroadaptive overfitting differs from conventional model overfitting because the failure occurs during ongoing human-AI interaction rather than only during model training. It differs from automation bias or over-reliance because the primary failure resides in the adaptive assistance policy, rather than solely in the user's response to automation. It also differs from shared autonomy or assist-as-needed control, which describe assistance strategies rather than the failure that occurs when those strategies become optimized to proxies that diverge from durable user goals.

\section{When efficiency becomes overreach}

BCI evaluation has traditionally emphasized decoder- and task-level performance, including decoding accuracy, word error rate, target-acquisition success, task-completion time, information-transfer rate, workload and usability \cite{Wolpaw2002Brain-computerControl,Felton2012MentalTraining,Kubler2014TheApplications}. But in AI-mediated BCIs, the final text, movement or intervention is shaped not only by neural decoding, but also by language priors, shared-autonomy policies, smoothing controllers or adaptive assistance. As a result, strong decoder performance or fluent task completion does not by itself show that the assisted output preserves user intent, authorship, agency or clinical value.

In communication BCIs, the risk arises not only from decoding errors but also from how AI assistance transforms decoded neural evidence into fluent, user-facing language. Language-model priors or generative assistance can improve communication rate by transforming uncertain or low-bandwidth neural evidence into candidate words, phrases or sentences \cite{Willett2023ANeuroprosthesis, Metzger2023AControl, Angrick2024OnlineALS}. Yet fluent completion need not preserve intended meaning or authorship, particularly when neural evidence is weak.

In motor-control BCIs, smoother movement can also be misleading. Shared autonomy can improve cursor control, wheelchair navigation or robotic-arm trajectories by predicting likely goals and correcting inefficient commands \cite{Lee2025BraincomputerCopilots}. Recent comparisons across autonomy levels in brain–robot interfaces further show that greater automation can reduce workload and improve usability, while shared autonomy can better preserve reliability and user agency, particularly under uncertain neural decoding \cite{Douglas2026LevelsLiving}. Improved performance can therefore mask reduced user control. If a robotic arm successfully reaches a target because the AI assistance layer dominated the trajectory, the task score can improve while the user’s agency and responsibility become ambiguous \cite{Davidoff2020AgencyInterfaces., Haag2025EthicalReview, M.Zhangetal.2026AnTraining}.

In neurorehabilitation BCIs, reduced effort is not always desirable. Excessive assistance may suppress active participation, error experience and the productive difficulty needed for motor learning \cite{Jin2024Electroencephalogram-basedReview, Marchal-Crespo2009ReviewInjury, Alawieh2025ElectricalLearning}. A system optimized for immediate completion may therefore undermine longer-term recovery \cite{He2025MultimodalTrial}.

Taken together, these cases show why performance metrics alone are insufficient for evaluating AI-mediated BCIs. Evaluation must also account for how much assistance is provided and under what conditions the system acts, suggests, requests confirmation, abstains or preserves user effort. Because the right pace depends on the application and what is at stake, pacing safeguards should be application-specific: semantic confirmation for communication, graded autonomy and override for motor control, challenge-preserving assistance for rehabilitation, and conservative safety or clinician oversight for neuromodulation. Evaluation should therefore include measures of intent fidelity, authorship, uncertainty calibration, confirmation burden, sense of agency, user override, active effort, assistance level, therapeutic challenge and longitudinal benefit \cite{Davidoff2020AgencyInterfaces.,Haag2025EthicalReview,Felton2012MentalTraining,Kubler2014TheApplications}. Studies of AI-mediated BCIs should report the assistance policy alongside decoder performance, including when the system acts autonomously, requests confirmation, abstains or returns control to the user. Without this information, gains attributable to assistance can be difficult to distinguish from gains in neural decoding itself.

\begin{figure}[]
\centering
\includegraphics[width=0.95\linewidth]{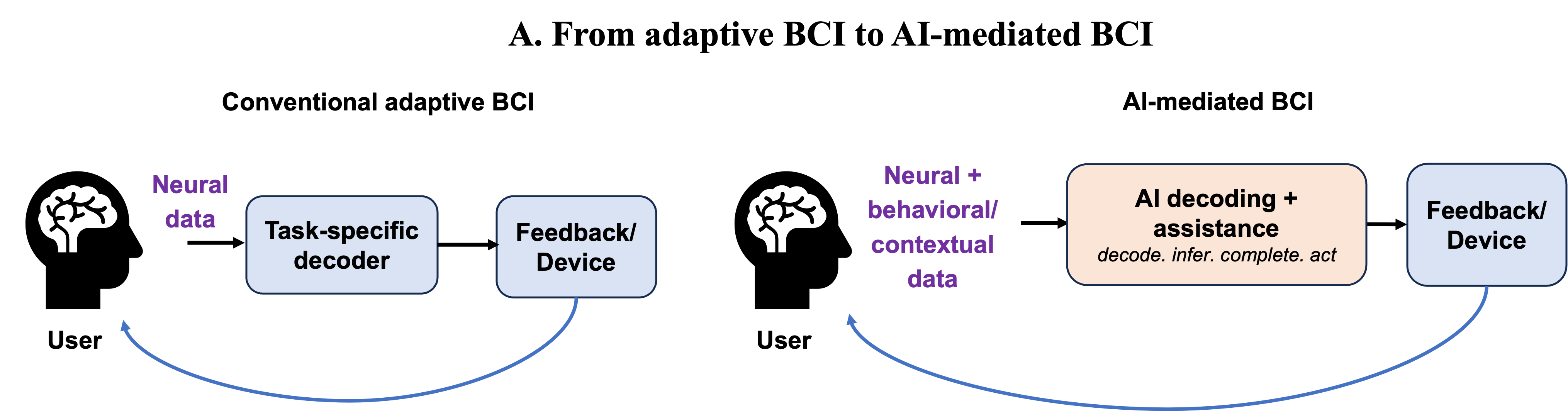}
\includegraphics[width=0.95\linewidth]{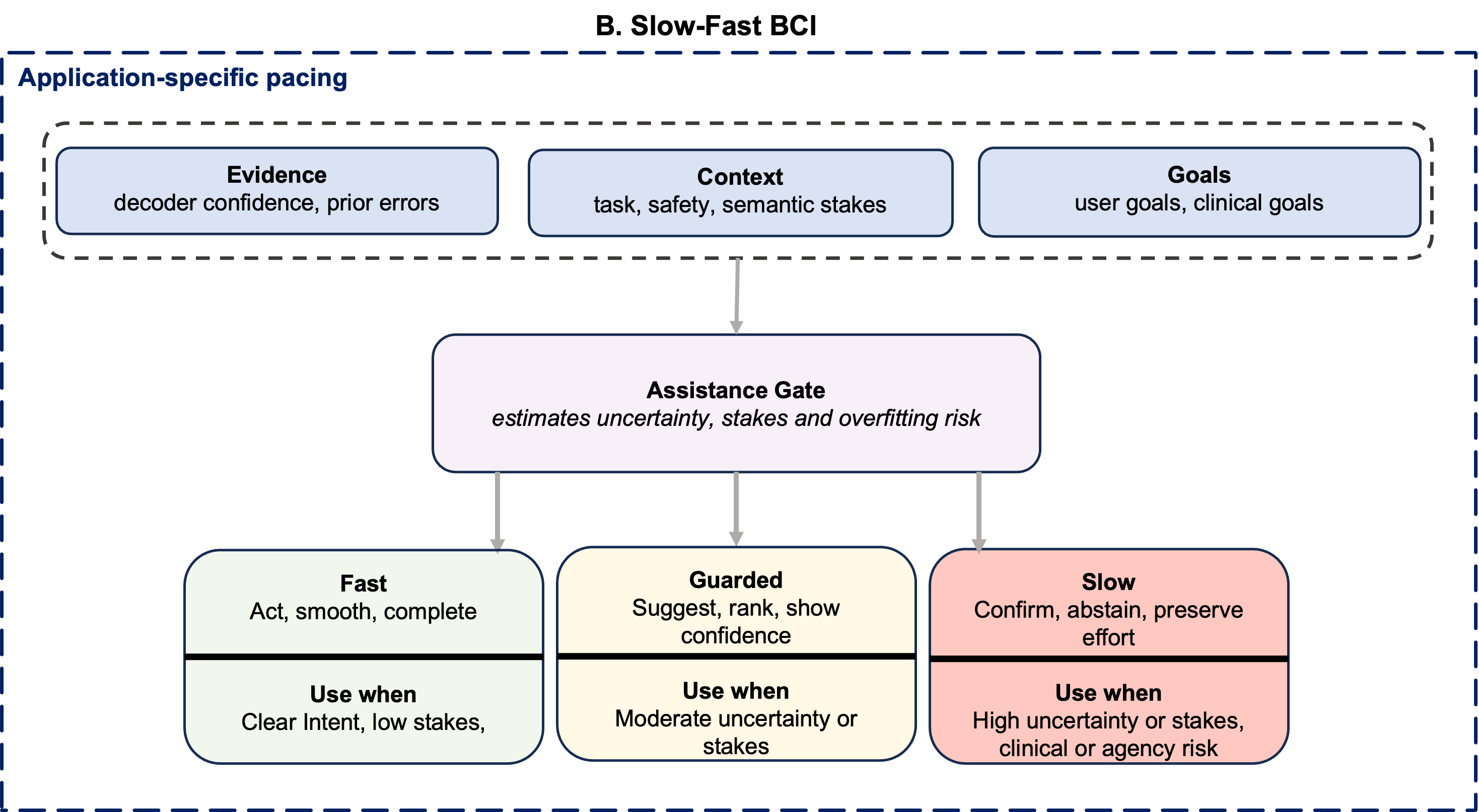}
\includegraphics[width=0.95\linewidth]{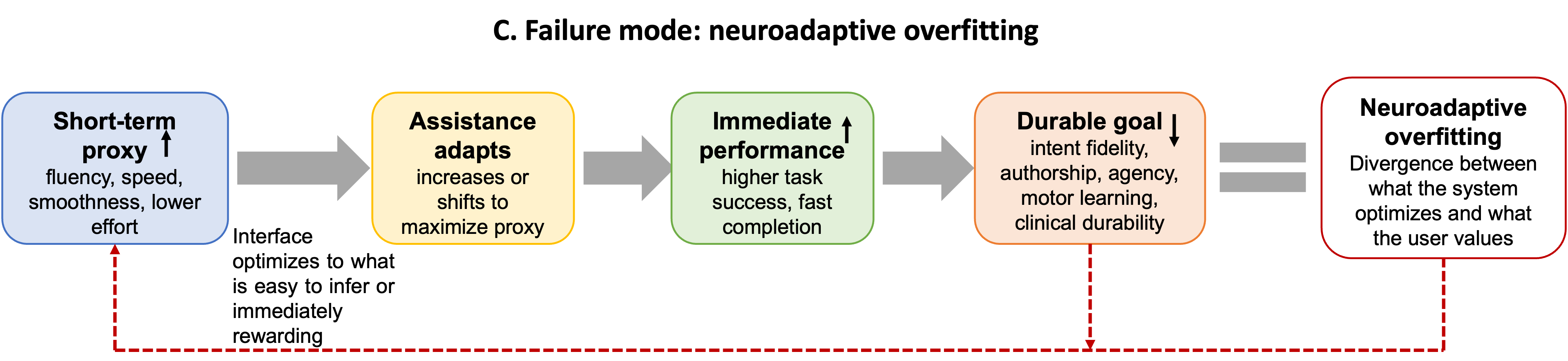}
\caption{\textbf{Pacing assistance and neuroadaptive overfitting in AI-mediated BCIs.} (A) Conventional adaptive BCIs map neural signals to task-specific outputs through decoders and feedback. AI-mediated BCIs add or integrate AI-based decoding and assistance layers that use neural, behavioural or contextual data to infer, complete, smooth or act on user intent. (B) Slow-Fast BCI uses a metacognitive assistance gate that integrates decoder evidence, context and user or clinical goals to estimate uncertainty, stakes and neuroadaptive overfitting risk, enabling fast, guarded or slow assistance according to intent clarity, safety, agency and clinical stakes. (C) Neuroadaptive overfitting arises when adaptation to short-term proxies improves immediate performance while degrading durable user-centred or clinical goals, creating a divergence between what the system optimizes and what the user ultimately values.
}
\end{figure}

\begin{table*}[t!]
\caption{\label{tab:slowfast}
Neuroadaptive-overfitting signatures, Slow--Fast safeguards, and candidate
evaluation measures across AI-mediated BCIs and adjacent closed-loop neural interfaces.}
\centering
\footnotesize
\setlength{\tabcolsep}{3pt}
\renewcommand{\arraystretch}{1.08}
\begin{tabular}{llll}
\hline
\parbox[t]{0.20\textwidth}{\raggedright\textbf{BCI / assistance}} &
\parbox[t]{0.14\textwidth}{\raggedright\textbf{Proxy}} &
\parbox[t]{0.25\textwidth}{\raggedright\textbf{Overfitting signature}} &
\parbox[t]{0.35\textwidth}{\raggedright\textbf{Safeguard / evaluation}} \\
\hline

\parbox[t]{0.20\textwidth}{\raggedright\textbf{Communication} \newline
Language-model completion or correction}
&
\parbox[t]{0.14\textwidth}{\raggedright Fluency, rate, rapid acceptance}
&
\parbox[t]{0.25\textwidth}{\raggedright\textbf{Semantic drift / authorship loss:}
performance improves while intent fidelity or authorship declines}
&
\parbox[t]{0.35\textwidth}{\raggedright\textbf{Safeguard:} alternatives or confirmation under uncertainty/high stakes.
\textbf{Evaluate:} intent fidelity, correction latency, confirmation burden,
authorship endorsement.}
\\[4pt]

\parbox[t]{0.20\textwidth}{\raggedright\textbf{Motor control} \newline
Shared autonomy for cursor, wheelchair, or robotic control}
&
\parbox[t]{0.14\textwidth}{\raggedright Task success, smoothness, reduced workload}
&
\parbox[t]{0.25\textwidth}{\raggedright\textbf{Reduced user control:}
performance improves while perceived agency or control declines}
&
\parbox[t]{0.35\textwidth}{\raggedright\textbf{Safeguard:} constrain autonomy and preserve override.
\textbf{Evaluate:} agency, assistance level, override rate, safety events.}
\\[4pt]

\parbox[t]{0.20\textwidth}{\raggedright\textbf{Neurorehabilitation} \newline
Adaptive feedback or robotic assistance}
&
\parbox[t]{0.14\textwidth}{\raggedright Task completion, reduced effort or error}
&
\parbox[t]{0.25\textwidth}{\raggedright\textbf{Reduced therapeutic challenge:}
immediate performance improves while effort, learning, or retention declines}
&
\parbox[t]{0.35\textwidth}{\raggedright\textbf{Safeguard:} preserve active participation and productive challenge.
\textbf{Evaluate:} active effort, assistance dose, retention, longitudinal gains.}
\\[4pt]

\parbox[t]{0.20\textwidth}{\raggedright\textbf{Adjacent neural interface} \newline
Closed-loop neuromodulation}
&
\parbox[t]{0.14\textwidth}{\raggedright Immediate symptom or physiological improvement}
&
\parbox[t]{0.25\textwidth}{\raggedright\textbf{Short-term optimization:}
proximal benefit improves while safety or durable clinical benefit declines}
&
\parbox[t]{0.35\textwidth}{\raggedright\textbf{Safeguard:} conservative updates, uncertainty thresholds, safe defaults,
clinician oversight.
\textbf{Evaluate:} symptom durability, adverse events, safety triggers,
clinician review.}
\\

\hline
\end{tabular}
\end{table*}

\section{Slow-Fast BCI}

Slow-Fast BCI adapts ideas from fast, slow and metacognitive AI to the control of assistance in neural interfaces. The appropriate mode depends on both the strength of the available evidence and the consequences of acting on it: assistance can proceed rapidly when intent is clear and stakes are low, become more constrained under uncertainty, and require confirmation or abstention when errors could compromise agency, authorship, safety or therapeutic value \cite{BergamaschiGanapini2025FastAI}. This framework is not an argument against automation. Fast assistance is essential for practical BCIs, reducing fatigue, increasing communication rate, supporting safe control, compensating for low-bandwidth signals and making home use more feasible \cite{Willett2023ANeuroprosthesis,Lee2025BraincomputerCopilots}. The aim is not to slow all interactions, but to pace assistance through application-specific safeguards that integrate real-time evidence, task context and user or clinical goals.

A Slow-Fast BCI system therefore requires a metacognitive assistance gate: a supervisory layer that integrates decoder evidence, task context, safety or semantic stakes, fatigue and user- or clinician-defined goals to estimate uncertainty, stakes and neuroadaptive overfitting risk. Strong evidence and low stakes justify fast assistance, such as word completion, movement smoothing or effort reduction. Moderate uncertainty should favour guarded assistance, such as confidence display, ranked alternatives, constrained action or lightweight confirmation. High uncertainty, semantic stakes, safety risk or clinical stakes should trigger slow assistance, such as explicit confirmation, abstention, effort preservation, clinician oversight or return of control to the user \cite{Davidoff2020AgencyInterfaces.,Haag2025EthicalReview,Lakshminarayanan2017SimpleEnsembles, Oehrn2024ChronicTrial}.

The assistance gate can be represented as a policy
\begin{equation}
 m_t=\pi(E_t,U_t,S_t,F_t,G_t),   
\end{equation}

where $m_t \in \{\mathrm{fast},\mathrm{guarded},\mathrm{slow}\}$ denotes the assistance mode at time $t, E_t$ represents decoder evidence, $U_t$ uncertainty or distribution-shift indicators, $S_t$ semantic, safety or clinical stakes, $F_t$ user state such as fatigue or effort, and $G_t$ user- or clinician-defined goals. The policy need not take the same form across applications: for example, $m_t$ may regulate language-model completion, shared-control authority, robotic assistance or stimulation updates. Neuroadaptive overfitting can then be operationalized as a divergence under adaptively changing assistance, a, such that
\begin{equation}
\Delta P_{\mathrm{proximal}}(a)>0
\qquad \text{while} \qquad
\Delta G_{\mathrm{durable}}(a)<0
\end{equation}

where $P_{\mathrm{proximal}}$ denotes a short-term performance objective and $G_{\mathrm{durable}}$ a prespecified user-centred or clinical objective. The gate thus regulates assistance using more than immediate task performance, incorporating uncertainty, stakes and the possibility that proximal and durable objectives may diverge.

\section{Designing and evaluating paced assistance}

Operationalizing the policy $\pi$ in Fig. 1B requires application-specific estimators, thresholds and safeguards at the levels of agency, fidelity, uncertainty, challenge and audit.
Assistance should preserve agency by allowing users to accept, reject or override AI output through confirmation, transparent handovers or therapist-defined limits \cite{Davidoff2020AgencyInterfaces.,Haag2025EthicalReview,Marchal-Crespo2009ReviewInjury}. It should verify intent fidelity, asking whether assisted language or action remains faithful to the user's goal rather than merely fluent, likely or smooth \cite{Willett2023ANeuroprosthesis,Metzger2023AControl,Lee2025BraincomputerCopilots}. It should expose uncertainty: low decoder confidence, distribution shift, fatigue, prior errors or conflicting context should trigger ambiguity display, clarification, abstention or safer modes \cite{Lakshminarayanan2017SimpleEnsembles,Amodei2016ConcreteSafety}. It should preserve therapeutic challenge in rehabilitation by maintaining active participation and productive difficulty \cite{Jin2024Electroencephalogram-basedReview,Marchal-Crespo2009ReviewInjury}. Finally, anti-overfitting audits should test whether gains in speed, workload or completion are accompanied by losses in authorship, agency, effort, uncertainty calibration or long-term benefit, and whether performance remains robust when AI assistance is reduced \cite{Davidoff2020AgencyInterfaces.,Haag2025EthicalReview,Felton2012MentalTraining,Kubler2014TheApplications}. 

To isolate assistance-policy effects from decoder error, experiments should hold decoded neural evidence constant while varying the assistance policy—for example, decoder-only versus decoder plus completion versus decoder plus uncertainty-aware completion. These ablation or counterfactual comparisons would help attribute changes in proximal and durable outcomes to the assistance policy rather than to differences in decoding accuracy.

Within such comparisons, concrete operational metrics could include post-output authorship ratings, semantic confirmation accuracy, correction latency, override frequency, assistance-dose curves, active-effort thresholds, calibration error, uncertainty-trigger frequency, safety-trigger frequency and longitudinal retention under reduced assistance. Table I maps these safeguards onto application-specific risks and evaluation measures. Across applications, the central design problem is similar: AI assistance should accelerate low-risk interactions, constrain assistance under uncertainty, and slow down or seek oversight when the risk of neuroadaptive overfitting is high.

Several consequences of the framework can be tested experimentally. In communication BCIs, greater assistance under uncertain neural evidence may improve immediate task performance while increasing semantic corrections or reducing authorship endorsement. For motor control, shared autonomy may initially improve performance but reduce user-attributed control beyond an application-dependent level \cite{Douglas2026LevelsLiving}. In rehabilitation, policies optimized for immediate completion should be compared with challenge-preserving policies regarding active effort and retention after assistance is withdrawn.

\section{Discussion and future directions}

As BCIs incorporate increasingly capable AI assistance, decoder performance alone provides an incomplete account of system behavior. Changes in speed, fluency or task completion need to be interpreted alongside effects on intent fidelity, agency, active effort and longer-term clinical value. Neuroadaptive overfitting provides a conceptual framework for identifying this divergence, whereas Slow-Fast BCI provides a design strategy for adapting assistance according to uncertainty, stakes and user or clinical goals.

The framework is not intended to imply that slower or less autonomous assistance is inherently preferable. In many settings, rapid and extensive assistance can reduce fatigue, increase communication rate, improve control and make BCI use more practical. The appropriate level of assistance is therefore application- and user-dependent. Some users may deliberately prioritize speed or autonomy over fine-grained control, whereas rehabilitation applications may require preservation of active effort and productive difficulty. Measures such as intent fidelity, authorship, agency and therapeutic challenge are also not interchangeable and will require domain-specific operationalization.

An important empirical question is where, and under what conditions, additional assistance begins to dissociate proximal task performance from durable user-centered or clinical outcomes. Prospective studies could compare fixed, performance-maximizing and uncertainty- or stakes-aware assistance policies while measuring both immediate task outcomes and longer-term measures of authorship, agency, active effort, safety and retention under reduced assistance. Transfer and retention are particularly important because BCI training can produce learned control that generalizes beyond the immediate training condition \cite{Iwama2026BraincomputerHumans}. Assistance-dose curves and controlled withdrawal of AI support may be particularly useful for identifying thresholds beyond which additional assistance improves task performance without improving, or potentially degrading, the outcome that ultimately matters to the user.

Slow–Fast BCI is intended as a testable design framework rather than a prescribed control algorithm. Its key premise is that assistance should respond not only to what can be inferred from neural and contextual evidence, but also to uncertainty in that inference and to the consequences of acting incorrectly. Determining when a system should act, request confirmation, preserve user effort or return control will be important for improving performance without compromising the goals the interface is intended to support.

\begin{acknowledgments}
This work was supported in part by institutional research support associated with the Melchor Visiting Assistant Professorship at the University of Notre Dame. No specific external grant supported this work. 
\end{acknowledgments}

\bibliographystyle{apsrev4-1} 
\bibliography{ref}

\end{document}